\documentclass[final,3p,times]{elsarticle}
\usepackage{amscd}
\usepackage{amssymb}
\usepackage{amsmath}
\usepackage{float}
\usepackage[T1]{fontenc}
\usepackage[latin1]{inputenc}
\usepackage{graphicx}
\usepackage{graphics}
\usepackage{color}
\usepackage{undertilde}
\def\be{\begin{equation}}
\def\ee{\end{equation}}
\def\ba{\begin{eqnarray}}
\def\ea{\end{eqnarray}}
\def\>{\rangle}
\def\<{\langle}
\def\n{\nonumber}

\journal{Physics Letters A}

\begin{document}

\begin{frontmatter}

\title{Field-induced dynamics in the quantum Brownian oscillator: An exact treatment}

\author{Ilki Kim}

\address{Department of Physics, North Carolina A$\&$T State
         University, Greensboro, North Carolina 27411, USA\\
         email) hannibal.ikim@gmail.com}

\begin{abstract}
We consider a quantum linear oscillator coupled to a bath in
equilibrium at an arbitrary temperature and then exposed to an
external field arbitrary in form and strength. We then derive the
reduced density operator in closed form of the coupled oscillator in
a non-equilibrium state at an arbitrary time.
\end{abstract}


\end{frontmatter}

\section{Introduction}\label{sec:introduction}
One of the most successful approaches to non-equilibrium statistical
mechanics is the linear response theory \cite{KUB57,KUB86,MAR08}.
This allows us to predict the average response of a physical
quantity of the system to external perturbations with weak strength.
At the heart of linear response theory we have the
fluctuation-dissipation theorem \cite{CAL51,WEI08}, which offers a
clear-cut relationship between irreversible processes in a
non-equilibrium state and thermal fluctuations in the (initial)
equilibrium state. However, this approach is, in general, restricted
to non-equilibrium states near equilibrium in validity.

As a prototype of quantum dissipative systems, the scheme of quantum
Brownian motion has been studied deeply and widely over a long
period \cite{WEI08}. At its heart we have a quantum harmonic
oscillator linearly coupled to an independent-oscillator model of a
heat bath [quantum Brownian oscillator] in equilibrium at a (low)
temperature. Due to its mathematical simplicity, this system allows
the linear response theory to yield an exact expression for an
average response of the system operator $\hat{q}$ (position) [and
also that of $\hat{p}$ (momentum)] to external forces $F$ arbitrary
in form and strength as well as those for the equilibrium
fluctuations $\<\hat{q}^2\>_{\beta}$ and $\<\hat{p}^2\>_{\beta}$,
respectively \cite{WEI08,ING98}. Based on this well-known result,
quantum Brownian oscillator has recently attracted considerable
interest in investigating thermodynamic behaviors of small-scaled
quantum systems coupled to quantum environments in the
low-temperature regime
\cite{ALL00,HAE05,FOR06,KIM06,KIM07,BUE08,LUT08,KIM10} (``quantum
thermodynamics'' \cite{SHE02,MAH04,CAP05}). Here the finite coupling
strength between system and environment yields some quantum
subtleties and so cannot be neglected whereas ordinary quantum
statistical mechanics is intrinsically based on a vanishingly small
coupling between them.

On the other hand, response functions in a far-from-equilibrium
state such as $\<\hat{q}^n(t)\>$ and $\<\hat{p}^n(t)\>$ with $n \geq
2$ in this system cannot be obtained directly from the linear
response theory. To explicitly have such non-equilibrium quantities,
we need to exactly treat the higher-order terms in the external
field and accordingly go beyond the scheme of linear response theory
\cite{ILK10,CUG07}. The primary goal of this paper is to derive a
reduced density operator in closed form of the coupled oscillator in
a non-equilibrium state at an arbitrary time $t$, which can,
obviously, provide all higher-order fluctuations of the
non-equilibrium state. For numerical analysis we will also consider
a variety of external fields (d.c. and a.c.) leading to explicit
evaluation of the non-equilibrium fluctuations. In doing so, we will
employ not only the $\hat{q} F(t)$ interaction Hamiltonian
(``scalar-potential gauge'') but also the $\hat{p} A(t)$ interaction
Hamiltonian (``vector-potential gauge''). The equivalence of the two
interactions is based on the gauge transformation between their
wavefunctions satisfying the corresponding (time-dependent)
Schr\"{o}dinger equations, respectively (for a detailed discussion
of $\hat{q} F$ versus $\hat{p} A$ gauge problem, see Ref.
\cite{SCH84}). And we will adopt the Drude model with a finite
frequency cut-off for the spectral density of bath modes, which is a
prototype for physically realistic damping \cite{WEI08}.

The general layout of this paper is the following. In Sect.
\ref{sec:basics} we briefly review the general results of quantum
Brownian oscillator needed for our later discussions. In Sect.
\ref{sec:density_operator} we explicitly derive an exact expression
for the reduced density operator of the coupled oscillator in a
non-equilibrium state. In Sect. \ref{sec:numerical} we perform
numerical analysis for various non-equilibrium quantities induced
from the reduced density operator. Finally we give the concluding
remarks of this paper in Sect. \ref{sec:conclusions}.

\section{General treatment of quantum Brownian oscillator}\label{sec:basics}
The quantum Brownian oscillator under consideration is described by
the model Hamiltonian \cite{WEI08,ING98}
\begin{equation}\label{eq:total_hamiltonian1}
    \hat{H}\; =\; \hat{H}_s\; +\; \hat{H}_{b-sb}\,,
\end{equation}
where
\begin{eqnarray}\label{eq:total_hamiltonian2}
    \hspace*{-.5cm}\hat{H}_s &=& \frac{\hat{p}^2}{2 M}\, +\,
    \frac{M}{2} \omega_0^2\,\hat{q}^2\\
    \hspace*{-.5cm}\hat{H}_{s-sb} &=& \sum_{j=1}^N \left\{\frac{\hat{p}_j^2}{2 m_j} +
    \frac{m_j}{2} \omega_j^2 \left(\hat{x}_j - \frac{c_j}{m_j\,\omega_j^2}\,\hat{q}\right)^2\right\}\,.
\end{eqnarray}
The Hamiltonian $\hat{H}_{s-sb}$ can split into the bath and the
coupling terms such as
\begin{eqnarray}
    \hat{H}_b &=& \sum_{j=1}^N \left(\frac{\hat{p}_j^2}{2 m_j} +
    \frac{m_j}{2} \omega_j^2\,\hat{x}_j^2\right)\\
    \hat{H}_{sb} &=& -\hat{q} \sum_{j=1}^N c_j\,\hat{x}_j\, +\, \hat{q}^2
    \sum_{j=1}^N \frac{c_j^2}{2 m_j\,\omega_j^2}\,,
\end{eqnarray}
respectively. The interaction Hamiltonian $\hat{H}_{sb}$ contains a
term of quadrature completion modifying the frequency of the system
oscillator $\hat{H}_s$. This quadrature term is needed not only for
later exact calculations; in fact, the model Hamiltonian without the
quadrature completion has a significant defect, as was discussed in
\cite{FOR88}. The total system is assumed to be within the canonical
thermal equilibrium state $\hat{\rho}_{\beta} = e^{-\beta
\hat{H}}/Z_{\beta}$ where $\beta = 1/(k_B\,T)$, and $Z_{\beta}$ is
the partition function.

From the Heisenberg equations of motion for $\hat{q}$ and $\hat{p}$
we can derive the quantum Langevin equation \cite{WEI08,ING98}
\begin{equation}\label{eq:eq_of_motion1}
    \textstyle M\,\ddot{\hat{q}}(t)\, +\,
    M \int_0^t d\tau\,\gamma(t-\tau)\,\dot{\hat{q}}(\tau)\,
    +\, M\,\omega_0^2\,\hat{q}(t)\; =\; \hat{\xi}(t)\,,
\end{equation}
where we used $\hat{p}(t) = M \dot{\hat{q}}(t)$, and the damping
kernel and the noise operator are, respectively, given as
\begin{equation}\label{eq:damping_kernel1}
    \gamma(t)\, =\, \frac{1}{M} \sum_{j=1}^N
    {\textstyle\frac{c_j^2}{m_j\,\omega_j^2}\cos(\omega_j\,t)\;\; ;\;\;
    \hat{\xi}(t)\, =\, -M \gamma(t)\,\hat{q}(0)}\, +\,
    \sum_{j=1}^N \textstyle c_j \left\{\hat{x}_j(0) \cos(\omega_j\,t)\, +\,
    \frac{\hat{p}_j(0)}{m_j\,\omega_j} \sin(\omega_j\,t)\right\}\,.
\end{equation}
Here the expectation value of the noise operator vanishes,
$\text{Tr}\,\{\hat{\xi}(t) \hat{\rho}_{\beta}\} = 0$ or,
equivalently, $\<\hat{\xi}(t)\>_{\rho_{b'}} = 0$ with respect to the
initial bath state, prepared as a shifted canonical equilibrium
distribution, $\hat{\rho}_{b'} = e^{-\beta
\hat{H}_{b-sb}}/Z^{(b')}_{\beta}$ in which a normalization constant
$Z^{(b')}_{\beta}$ is the properly defined partition function
\cite{WEI08}. And the noise correlation is given as \cite{HAE05}
\begin{equation}\label{eq:qm_noise_correlation1}
    S_{\xi\xi}(t-t')\, :=\, \frac{1}{2}
    \left\<\hat{\xi}(t)\,\hat{\xi}(t')\,+\,\hat{\xi}(t')\,\hat{\xi}(t)\right\>_{\rho_{b'}}\,
    =\, \frac{\hbar}{2} \sum_{j=1}^N \textstyle\frac{c_j^2}{m_j\,\omega_j}
    \cos\{\omega_j (t-t')\}\,\coth\left(\frac{\beta \hbar \omega_j}{2}\right)\,.
\end{equation}
With $\hbar \to 0$, the correlation $S_{\xi\xi}(t-t')$ reduces to
its classical counterpart, $M \gamma(t-t')/\beta$ \cite{ILK07}. We
also introduce a response function \cite{ING98}
\begin{equation}\label{eq:response_fkt1}
    \textstyle\chi_{qq}(t)\, =\, \frac{i}{\hbar}\,\left\<[\hat{q}(t),
    \hat{q}(0)]\right\>_{\beta}\, \Theta(t)\,,
\end{equation}
where $\Theta(t)$ represents a step function. Then we can have other
response functions as well such as $\chi_{pq}(t) = -\chi_{qp}(t) = M
\dot{\chi}_{qq}(t)$, and $\chi_{pp}(t) = -M^2 \ddot{\chi}_{qq}(t)$.

For a later purpose it is instructive to discuss the time-reversal
dynamics of $\hat{q}(t)$ in terms of $\hat{r}(t) := \hat{q}(-t)$ and
its momentum $\hat{s}(t) := -\hat{p}(-t)$. We can then derive the
corresponding quantum Langevin equation \cite{HAE05}
\begin{equation}\label{eq:eq_of_motion1_for_x}
    \textstyle M\,\ddot{\hat{r}}(t)\, +\,
    M \int_0^t d\tau\,\gamma(t-\tau)\,\dot{\hat{r}}(\tau)\,
    +\, M\,\omega_0^2\,\hat{r}(t)\; =\; \hat{\xi}_{-}(t)\,.
\end{equation}
While this is the same in form as equation (\ref{eq:eq_of_motion1}),
the two equations differ in the noise in such a way that
$\hat{\xi}_{-}(t)$ is identical to $\hat{\xi}(t)$, however, with the
replacement of $\hat{p}_j(0) \to -\hat{p}_j(0)$ in
(\ref{eq:damping_kernel1}). And from equation
(\ref{eq:response_fkt1}) and stationarity of the equilibrium
correlation function between operators $\hat{F}$ and $\hat{G}$ such
as $\<\hat{F}(t)\,\hat{G}(0)\>_{\beta} =
\<\hat{F}(0)\,\hat{G}(-t)\>_{\beta}$ \cite{ING98}, we can easily
obtain $\chi_{rr}(t) = -\chi_{qq}(t)$ [note that $\hat{r}(0) =
\hat{q}(0)$]. Likewise, it also appears that $\chi_{rp}(t) =
\chi_{qp}(t)$.

Now we intend to derive explicit expressions for $\hat{q}(t)$ and
$\hat{r}(t)$, respectively. To do so, we first apply the Laplace
transforms to equations (\ref{eq:eq_of_motion1}) and
(\ref{eq:eq_of_motion1_for_x}), respectively. Let the Laplace
transform $\underline{\hat{q}}(s) := {\mathcal L}\{\hat{q}(t)\}(s)$,
and so we have ${\mathcal L}\{\dot{\hat{q}}(t)\}(s) = s
\underline{\hat{q}}(s) - \hat{q}(0)$ and ${\mathcal
L}\{\ddot{\hat{q}}(t)\}(s) = s^2 \underline{\hat{q}}(s) - s
\hat{q}(0) - \dot{\hat{q}}(0)$ \cite{ROB66}. We can then obtain the
Fourier-Laplace transform of $\hat{q}(t)$, which reads as
\begin{equation}
    \utilde{\hat{q}}(\omega)\, =\, {\textstyle\frac{i \omega}{\omega^2 + i \omega \tilde{\gamma}(\omega) -
    \omega_0^2}\,\hat{q}\,-\,\frac{1}{M} \frac{1}{\omega^2 + i \omega \tilde{\gamma}(\omega) -
    \omega_0^2}\,\hat{p}\,-\,\frac{1}{M} \frac{1}{\omega^2 + i \omega \tilde{\gamma}(\omega) -
    \omega_0^2}} \sum_j \textstyle\frac{c_j}{\omega^2 - \omega_j^2} \left(i \omega \hat{x}_j -
    \frac{\hat{p}_j}{m_j}\right)\label{eq:laplace2_1}
\end{equation}
where $\utilde{\hat{q}}(\omega) = \underline{\hat{q}}(s)$ with $s =
-i \omega + 0^+$, and the Fourier-Laplace transform of $\hat{r}(t)$,
\begin{equation}
    \utilde{\hat{r}}(\omega)\, =\, {\textstyle\frac{i \omega}{\omega^2 + i \omega \tilde{\gamma}(\omega) -
    \omega_0^2}\,\hat{q}\,+\,\frac{1}{M} \frac{1}{\omega^2 + i \omega \tilde{\gamma}(\omega) -
    \omega_0^2}\,\hat{p}\,-\,\frac{1}{M} \frac{1}{\omega^2 + i \omega \tilde{\gamma}(\omega) -
    \omega_0^2}} \sum_j \textstyle\frac{c_j}{\omega^2 - \omega_j^2} \left(i \omega \hat{x}_j +
    \frac{\hat{p}_j}{m_j}\right)\label{eq:laplace2_2}
\end{equation}
where $\utilde{\hat{r}}(\omega) = \underline{\hat{r}}(s)$. The
operators $\hat{q}, \hat{p}, \hat{x}_j$, and $\hat{p}_j$ represent
the initial values $\hat{q}(0), \hat{p}(0), \hat{x}_j(0)$, and
$\hat{p}_j(0)$, respectively. And the Fourier-Laplace transform of
$\gamma(t)$ is
\begin{equation}
    \tilde{\gamma}(\omega)\; =\; \frac{i \omega}{M} \sum_j^N \frac{c_j^2}{m_j\,\omega_j^2}\,
    \frac{1}{\omega^2 - \omega_j^2}\,.\label{eq:gamma_tilde1}
\end{equation}
We introduce the susceptibility, defined as the Fourier-Laplace
transform of $\chi_{qq}(t)$ in (\ref{eq:response_fkt1}), such as
\cite{WEI08,ING98}
\begin{equation}\label{eq:susceptibility1}
    \textstyle\tilde{\chi}_{qq}(\omega)\, :=\,
    \int_{-\infty}^{\infty} dt\,\chi_{qq}(t)\,e^{i \omega t}\, =\, \frac{i}{\hbar}\,\<[\utilde{\hat{q}}(\omega), \hat{q}]\>_{\beta}\,,
\end{equation}
which easily reduces to $-1/\{M (\omega^2 + i \omega
\tilde{\gamma}(\omega) - \omega_0^2)\}$ with the aid of
(\ref{eq:laplace2_1}). And it then appears that
\begin{eqnarray}\label{eq:susceptibility2_1}
    \hspace*{-.6cm}\textstyle\tilde{\chi}_{q x_j}(\omega) &=& \textstyle\frac{i}{\hbar}\,\<[\utilde{\hat{q}}(\omega),
    \hat{x}_j]\>_{\beta}\; =\; \frac{c_j}{m_j\,(\omega_j^2 - \omega^2)}\,\tilde{\chi}_{qq}(\omega)\,,\\
    \hspace*{-.6cm}\textstyle\tilde{\chi}_{rr}(\omega) &=& \textstyle\frac{i}{\hbar}\,\<[\utilde{\hat{r}}(\omega),
    \hat{q}]\>_{\beta}\; =\; -\tilde{\chi}_{qq}(\omega)\,,
\end{eqnarray}
$\tilde{\chi}_{qp}(\omega) = i \omega M \tilde{\chi}_{qq}(\omega)$,
and $\tilde{\chi}_{q p_j}(\omega) = i \omega m_j \tilde{\chi}_{q
x_j}(\omega)$. Likewise, we also have $\tilde{\chi}_{rp}(\omega) =
\tilde{\chi}_{qp}(\omega)$, $\tilde{\chi}_{r x_j}(\omega) =
-\tilde{\chi}_{q x_j}(\omega)$, and $\tilde{\chi}_{r p_j}(\omega) =
\tilde{\chi}_{q p_j}(\omega)$. Consequently, equations
(\ref{eq:laplace2_1}) and (\ref{eq:laplace2_2}) can be rewritten in
terms of the susceptibilities, respectively, as the compact
expressions
\begin{subequations}
\begin{eqnarray}
    \utilde{\hat{q}}(\omega) &=&
    {\textstyle-\tilde{\chi}_{qp}(\omega)\,\hat{q} +
    \tilde{\chi}_{qq}(\omega)\,\hat{p}} - \sum_{j} \textstyle\{\tilde{\chi}_{q p_j}(\omega)\,\hat{x}_j - \tilde{\chi}_{q x_j}(\omega)\,\hat{p}_j\}\label{eq:q_frequency1}\\
    \utilde{\hat{r}}(\omega) &=&
    {\textstyle-\tilde{\chi}_{rp}(\omega)\,\hat{q} +
    \tilde{\chi}_{rr}(\omega)\,\hat{p}} - \sum_{j} \textstyle\{\tilde{\chi}_{r p_j}(\omega)\,\hat{x}_j - \tilde{\chi}_{r x_j}(\omega)\,\hat{p}_j\}\,.\label{eq:r_frequency1}
\end{eqnarray}
\end{subequations}
Applying the inverse Fourier transforms to (\ref{eq:q_frequency1})
and (\ref{eq:r_frequency1}) immediately allow us to have
\begin{subequations}
\begin{eqnarray}
    \hat{q}(t) &=& {\textstyle-\chi_{qp}(t)\,\hat{q} + \chi_{qq}(t)\,\hat{p}} - \sum_{j} \textstyle\{\chi_{q p_j}(t)\,\hat{x}_j - \chi_{q x_j}(t)\,\hat{p}_j\}\label{eq:q_time1}\\
    \hat{r}(t) &=& {\textstyle-\chi_{rp}(t)\,\hat{q} + \chi_{rr}(t)\,\hat{p}} - \sum_{j} \textstyle\{\chi_{r p_j}(t)\,\hat{x}_j - \chi_{r x_j}(t)\,\hat{p}_j\}\,,\label{eq:r_time1}
\end{eqnarray}
\end{subequations}
where
\begin{equation}\label{eq:inverse_laplace}
    \textstyle\chi_{qq}(t)\, =\, \frac{1}{2 \pi} \int_{-\infty}^{\infty} d\omega\,\tilde{\chi}_{qq}(\omega)\,e^{-i \omega t}\,.
\end{equation}
The two equations will be used in Sect. \ref{sec:density_operator}
for derivation of the reduced density operator of the coupled
oscillator in closed form.

For a later purpose it is useful to introduce well-known expressions
for the equilibrium fluctuations in terms of the susceptibility
$\tilde{\chi}_{qq}(\omega)$ such as \cite{FOR85}
\begin{equation}\label{eq:x_correlation1}
    \textstyle\<\hat{q}^2\>_{\beta}\,=\,\textstyle\frac{\hbar}{\pi} \int_0^{\infty}
    d\omega\,\coth\left(\frac{\beta \hbar \omega}{2}\right)\,\text{Im}\{\tilde{\chi}_{qq}(\omega +
    i\,0^+)\}
\end{equation}
and
\begin{equation}\label{eq:x_dot_correlation1}
    \textstyle\<\hat{p}^2\>_{\beta}\,=\,\textstyle\frac{M^2 \hbar}{\pi} \int_0^{\infty}
    d\omega\,\omega^2\,\coth\left(\frac{\beta \hbar \omega}{2}\right)\,\text{Im}\{\tilde{\chi}_{qq}(\omega +
    i\,0^+)\}\,,
\end{equation}
respectively, which can be derived from the fluctuation-dissipation
theorem \cite{CAL51,WEI08}.

\section{Reduced density operator of the coupled oscillator in a non-equilibrium state}\label{sec:density_operator}
Now we study the influence of an external field on a linear
oscillator coupled to a bath. To do so, we first consider the
equation of motion for the density operator of the total system
({\em i.e.}, oscillator plus bath), which reads \cite{WEI08,ING98}
\begin{equation}\label{eq:density_fkt1}
    \textstyle\hat{\rho}(t)\; =\; e^{-i t \hat{\mathcal L}_0} \hat{\rho}(0)\,-\,i \int_0^t d\tau\,e^{-i (t - \tau) \hat{\mathcal{L}}_0}
    \hat{\mathcal L}_1(\tau)\,\hat{\rho}(\tau)\,.
\end{equation}
In the scalar-potential gauge for the field-coupling, the total
Hamiltonian reads as $\hat{{\mathcal H}}_s(t) = \hat{H} -
\hat{q}\,F(t)$, and the corresponding Liouville operator
$\hat{{\mathcal L}} = \hat{{\mathcal L}}_0 + \hat{{\mathcal
L}}_1^{(s)}$ satisfies
\begin{equation}\label{eq:density_fkt2}
    \textstyle\hat{{\mathcal L}}_0\,\hat{\rho}(t)\,=\,\frac{1}{\hbar}
    [\hat{H}, \hat{\rho}(t)]\,;\,\hat{{\mathcal
    L}}_1^{(s)}(\tau)\,\hat{\rho}(\tau)\,=\,-\frac{1}{\hbar} [\hat{q}, \hat{\rho}(\tau)]\,F(\tau)\,.
\end{equation}
In the vector-potential gauge, on the other hand, the total
Hamiltonian is given as
\begin{equation}\label{eq:vector_potential_hamiltonian1}
    \hat{{\mathcal H}}_v(t)\; =\; \frac{\left\{\hat{p}\,+\,p_c(t)\right\}^2}{2
    M}\,+\,\frac{M \omega_0^2}{2}\,\hat{q}^2\,+\,\hat{H}_{s-sb}\,,
\end{equation}
which is identical $\hat{H} + p_c(t)\,\hat{p}/M + p_c^2(t)/(2 M)$,
and accordingly
\begin{equation}\label{eq:density_fkt2_1}
    \textstyle\hat{{\mathcal L}}_0\,\hat{\rho}(t)\,=\,\frac{1}{\hbar}
    [\hat{H}, \hat{\rho}(t)]\;;\;\hat{{\mathcal
    L}}_1^{(v)}(\tau)\,\hat{\rho}(\tau)\,=\,\frac{1}{\hbar} [\hat{p}, \hat{\rho}(\tau)]\,\frac{p_c(\tau)}{M}\,.
\end{equation}
Here the vector potential $p_c(t) = \int_0^t F(\tau)\,d\tau$ is an
impulse induced by field $F$. For an uncoupled oscillator the
equivalence of the two interactions is based on the gauge
transformation $\psi_{s}(q,t) = e^{\frac{i}{\hbar} q\cdot p_c(t)}
\psi_v(q,t)$ \cite{SCH84}, where the wavefunctions $\psi_{s}(q,t)$
and $\psi_{s}(q,t)$ satisfy the time-dependent Schr\"{o}dinger
equations in the scalar-potential and the corresponding
vector-potential gauges, respectively.

We substitute (\ref{eq:density_fkt2}) into (\ref{eq:density_fkt1}),
with $\hat{\rho}(0) = \hat{\rho}_{\beta}$, and make iterations for
$\hat{\rho}(\tau)$ in the integral. Then we can arrive at the
expression
\begin{eqnarray}\label{eq:density_fkt5}
    \textstyle\hat{\rho}_s(t) &=& \textstyle\hat{\rho}_{\beta}\,+\,\frac{i}{\hbar} \int_0^t d\tau\,F(\tau)\,e^{-\frac{i}{\hbar} (t-\tau) \hat{H}}
    \left[\hat{q}, \hat{\rho}_{\beta}\right] e^{\frac{i}{\hbar} (t-\tau) \hat{H}}\,+\,\left(\frac{i}{\hbar}\right)^2
    \int_0^t d\tau\,F(\tau) \int_0^{\tau} d\tau'\,F(\tau')\,e^{-\frac{i}{\hbar} (t-\tau) \hat{H}} \times\n\\
    && \textstyle\left[\hat{q}, e^{-\frac{i}{\hbar} (\tau-\tau') \hat{H}}\left[\hat{q},
    \hat{\rho}_{\beta}\right] e^{\frac{i}{\hbar} (\tau-\tau') \hat{H}}\right] e^{\frac{i}{\hbar} (t-\tau)
    \hat{H}}\,+\,\left(\frac{i}{\hbar}\right)^3 \int_0^t d\tau\,F(\tau) \int_0^{\tau} d\tau'\,F(\tau') \int_0^{\tau'} d\tau''\,F(\tau'')\,e^{-\frac{i}{\hbar} (t-\tau) \hat{H}} \times\n\\
    && \textstyle\left[\hat{q}, e^{-\frac{i}{\hbar} (\tau-\tau') \hat{H}} \left[\hat{q},
    e^{-\frac{i}{\hbar} (\tau'-\tau'') \hat{H}} \left[\hat{q},
    \hat{\rho}_{\beta}\right] e^{\frac{i}{\hbar} (\tau'-\tau'') \hat{H}}\right]
    e^{\frac{i}{\hbar} (\tau-\tau') \hat{H}}\right] e^{\frac{i}{\hbar} (t-\tau) \hat{H}}\,+\,\cdots\,.
\end{eqnarray}
With the aid of $[\hat{\rho}_{\beta}, \hat{H}] = 0$, this equation
easily reduces to the expression in terms of $\hat{r}(t) =
e^{-\frac{i}{\hbar} t \hat{H}} \hat{q}\,e^{\frac{i}{\hbar} t
\hat{H}}$ such as
\begin{eqnarray}\label{eq:density_fkt6}
    \hat{\rho}_s(t) &=& \textstyle\hat{\rho}_{\beta}\,+\,\frac{i}{\hbar} \int_0^t d\tau\,F(\tau)\,\left[\hat{r}(t-\tau),
    \hat{\rho}_{\beta}\right]\,+\,\left(\frac{i}{\hbar}\right)^2 \int_0^t d\tau\,F(\tau) \int_0^{\tau} d\tau'\,F(\tau')\,\left[\hat{r}(t-\tau),
    \left[\hat{r}(t-\tau'), \hat{\rho}_{\beta}\right]\right]\,+\n\\
    && \textstyle\left(\frac{i}{\hbar}\right)^3
    \int_0^t d\tau\,F(\tau) \int_0^{\tau} d\tau'\,F(\tau') \int_0^{\tau'}
    d\tau''\,F(\tau'')\,\left[\hat{r}(t-\tau), \left[\hat{r}(t-\tau'), \left[\hat{r}(t-\tau''),
    \hat{\rho}_{\beta}\right]\right]\right]\,+\,\cdots\,,
\end{eqnarray}
where
\begin{equation}\label{eq:density_fkt7}
    \left[\hat{r}(t), \hat{\rho}_{\beta}\right]\, =\, {\textstyle-\chi_{rp}(t)
    \left[\hat{q}, \hat{\rho}_{\beta}\right]\,+\,\chi_{rr}(t) \left[\hat{p},
    \hat{\rho}_{\beta}\right]}\,-\,\sum_j {\textstyle\chi_{r p_j}(t) \left[\hat{x}_j,
    \hat{\rho}_{\beta}\right]}\,+\,\sum_j \textstyle\chi_{r x_j}(t) \left[\hat{p}_j,
    \hat{\rho}_{\beta}\right]\,,
\end{equation}
obtained directly from equation (\ref{eq:r_time1}). For the
vector-potential gauge, on the other hand, we plug
(\ref{eq:density_fkt2_1}) into (\ref{eq:density_fkt1}), and after
making some calculations similar to those for
(\ref{eq:density_fkt5}) we can finally obtain the expression in
terms of $\hat{s}(t) = -e^{-\frac{i}{\hbar} t \hat{H}}
\hat{p}\,e^{\frac{i}{\hbar} t \hat{H}}$ such as
\begin{eqnarray}\label{eq:density_fkt6_1}
    \hat{\rho}_v(t) &=& \textstyle\hat{\rho}_{\beta}\,+\,\frac{i}{\hbar M} \int_0^t d\tau\,p_c(\tau)\,\left[\hat{s}(t-\tau),
    \hat{\rho}_{\beta}\right]\,+\,\left(\frac{i}{\hbar M}\right)^2 \int_0^t d\tau\,p_c(\tau) \int_0^{\tau} d\tau'\,p_c(\tau')\,\left[\hat{s}(t-\tau),
    \left[\hat{s}(t-\tau'), \hat{\rho}_{\beta}\right]\right]\,+\n\\
    && \textstyle\left(\frac{i}{\hbar M}\right)^3
    \int_0^t d\tau\,p_c(\tau) \int_0^{\tau} d\tau'\,p_c(\tau') \int_0^{\tau'}
    d\tau''\,p_c(\tau'')\,\left[\hat{s}(t-\tau), \left[\hat{s}(t-\tau'), \left[\hat{s}(t-\tau''), \hat{\rho}_{\beta}\right]\right]\right]\,+\,\cdots\,.
\end{eqnarray}
From (\ref{eq:density_fkt7}) and $\hat{s}(t) = M \dot{\hat{r}}(t)$
we can easily have
\begin{equation}\label{eq:density_fkt7_1}
    \left[\hat{s}(t), \hat{\rho}_{\beta}\right]\, =\, {\textstyle-\chi_{sp}(t)
    \left[\hat{q}, \hat{\rho}_{\beta}\right]\,+\,\chi_{sr}(t) \left[\hat{p},
    \hat{\rho}_{\beta}\right]}\,-\,\sum_j {\textstyle\chi_{s p_j}(t) \left[\hat{x}_j,
    \hat{\rho}_{\beta}\right]}\,+\,\sum_j \textstyle\chi_{s x_j}(t) \left[\hat{p}_j,
    \hat{\rho}_{\beta}\right]\,.
\end{equation}

Let us now consider the reduced density operators for the coupled
oscillator, $\hat{{\mathcal R}}(t) := \text{Tr}_b\,\hat{\rho}(t)$
from (\ref{eq:density_fkt6}) and (\ref{eq:density_fkt6_1}),
respectively. Here, $\text{Tr}_b$ denotes the partial trace for the
bath alone. The initial state $\hat{{\mathcal R}}(0)$ of the coupled
oscillator, being the reduced operator of the canonical equilibrium
state $\hat{\rho}_{\beta}$, is known as \cite{WEI08,GRA88}
\begin{equation}\label{eq:density_operator1}
    \textstyle\<q|\hat{{\mathcal R}}(0)|q'\>\, =\, \frac{1}{\sqrt{2\pi \<\hat{q}^2\>_{\beta}}}\, \exp\left(-\frac{(q + q')^2}{8\,\<\hat{q}^2\>_{\beta}} -
    \frac{\<\hat{p}^2\>_{\beta}\,(q - q')^2}{2 \hbar^2}\right)\,.
\end{equation}
First, from (\ref{eq:density_fkt7}) we can obtain
\begin{equation}\label{eq:density_fkt8}
    \textstyle\<q|\text{Tr}_b \left[\hat{r}(t), \hat{\rho}_{\beta}\right]|q'\>\, =\, \hat{{\mathcal S}}_{qq'}(t)\,\<q|\hat{{\mathcal R}}(0)|q'\>\,,
\end{equation}
where $\hat{{\mathcal S}}_{qq'}(t) := -i
\hbar\,\chi_{rr}(t)\,\left(\partial_q + \partial_{q'}\right) -
\chi_{rp}(t)\,(q - q')$, and similarly from
(\ref{eq:density_fkt7_1}) we can also have
\begin{equation}\label{eq:density_fkt8_1}
    \textstyle\<q|\text{Tr}_b \left[\hat{s}(t), \hat{\rho}_{\beta}\right]|q'\>\, =\, \hat{{\mathcal V}}_{qq'}(t)\,\<q|\hat{{\mathcal R}}(0)|q'\>\,,
\end{equation}
where $\hat{{\mathcal V}}_{qq'}(t) := -i
\hbar\,\chi_{sr}(t)\,\left(\partial_q + \partial_{q'}\right) -
\chi_{sp}(t)\,(q - q')$. Here we used
\begin{subequations}
\begin{eqnarray}
    {\textstyle\int_{-\infty}^{\infty}}\prod_k dx_k\,\<x_k|
    \left[\hat{x}_j,\hat{\rho}_{\beta}\right] |x_k\> &=& 0\label{eq:density_fkt9}\\
    {\textstyle\int_{-\infty}^{\infty}}\prod_k dp_k\,\<p_k|
    \left[\hat{p}_j,\hat{\rho}_{\beta}\right] |p_k\> &=& 0\,.\label{eq:density_fkt10}
\end{eqnarray}
\end{subequations}
From (\ref{eq:r_time1}), (\ref{eq:density_fkt9}) and
(\ref{eq:density_fkt10}) it also appears that
\begin{equation}\label{eq:density_fkt10_1}
    {\textstyle\int_{-\infty}^{\infty}} \prod_k\textstyle dx_k \left\<x_k\left|
    \left[\hat{{\mathcal O}}, \left[\hat{{\mathcal O}}', \hat{\rho}_{\beta}\right]\right] \right|x_k\right\>\, =\, 0
\end{equation}
unless both $\hat{{\mathcal O}} \in \{\hat{q}, \hat{p}\}$ and
$\hat{{\mathcal O}}' \in \{\hat{q}, \hat{p}\}$. Therefore we can
arrive at the expressions,
\begin{eqnarray}
    \<q|\text{Tr}_b \left[\hat{r}(t),
    \left[\hat{r}(\tau), \hat{\rho}_{\beta}\right]\right]|q'\> &=&
    \textstyle\hat{{\mathcal S}}_{qq'}(t)\,\hat{{\mathcal S}}_{qq'}(\tau)\,\<q|\hat{{\mathcal R}}(0)|q'\>\label{eq:density_fkt13}\\
    \<q|\text{Tr}_b \left[\hat{s}(t),
    \left[\hat{s}(\tau), \hat{\rho}_{\beta}\right]\right]|q'\> &=&
    \textstyle\hat{{\mathcal V}}_{qq'}(t)\,\hat{{\mathcal V}}_{qq'}(\tau)\,\<q|\hat{{\mathcal R}}(0)|q'\>\,.\label{eq:density_fkt13_1}
\end{eqnarray}
Here we also used
\begin{eqnarray}\label{eq:density_fkt12}
    \hspace*{-1.3cm}&&\textstyle\<q|\text{Tr}_b \left[\hat{p}, \left[\hat{p},
    \hat{\rho}_{\beta}\right]\right]|q'\> =
    \textstyle\left(\frac{\hbar}{i}\right)^2 \left(\partial_q + \partial_{q'}\right)^2 \<q|\hat{{\mathcal R}}(0)|q'\>\n\\
    \hspace*{-1.3cm}&&\textstyle\<q|\text{Tr}_b \left[\hat{q}, \left[\hat{p},
    \hat{\rho}_{\beta}\right]\right]|q'\> =
    \frac{\hbar}{i} \left(q - q'\right) \left(\partial_q + \partial_{q'}\right) \<q|\hat{{\mathcal R}}(0)|q'\>\n\\
    \hspace*{-1.3cm}&&\textstyle\<q|\text{Tr}_b \left[\hat{q}, \left[\hat{q},
    \hat{\rho}_{\beta}\right]\right]|q'\> = \textstyle\left(q - q'\right)^2 \<q|\hat{{\mathcal R}}(0)|q'\>\,,
\end{eqnarray}
and $[\hat{q}, [\hat{p}, \hat{\rho}_{\beta}]] = [\hat{p}, [\hat{q},
\hat{\rho}_{\beta}]]$. Along the same line, after making lengthy
calculations, we can also obtain
\begin{eqnarray}
    \<q|\text{Tr}_b \left[\hat{r}(t), \left[\hat{r}(\tau),
    \left[\hat{r}(\tau'), \hat{\rho}_{\beta}\right]\right]\right]|q'\> &=& \hat{{\mathcal S}}_{qq'}(t)\,\hat{{\mathcal S}}_{qq'}(\tau)\,\hat{{\mathcal S}}_{qq'}(\tau')\,\<q|\hat{{\mathcal
    R}}(0)|q'\>\label{eq:density_fkt14}\\
    \<q|\text{Tr}_b \left[\hat{s}(t), \left[\hat{s}(\tau),
    \left[\hat{s}(\tau'), \hat{\rho}_{\beta}\right]\right]\right]|q'\> &=& \hat{{\mathcal V}}_{qq'}(t)\,\hat{{\mathcal V}}_{qq'}(\tau)\,\hat{{\mathcal V}}_{qq'}(\tau')\,\<q|\hat{{\mathcal
    R}}(0)|q'\>\,.\label{eq:density_fkt14_1}
\end{eqnarray}
With the help of equations (\ref{eq:density_fkt6}),
(\ref{eq:density_fkt8}), (\ref{eq:density_fkt13}), and
(\ref{eq:density_fkt14}) we can finally find the matrix elements of
the reduced density operator in the scalar-potential gauge such as
\begin{equation}\label{eq:density_fkt16}
    \<q|\hat{{\mathcal R}}_s(t)|q'\>\, =\, T\,e^{\frac{i}{\hbar} \hat{J}_s(t)}\,\<q|\hat{{\mathcal
    R}}(0)|q'\>\,,
\end{equation}
where the operator $\hat{J}_s(t) := \int_0^t
d\tau\,F(\tau)\,\hat{{\mathcal S}}_{qq'}(t-\tau)$ represents a
time-evolution action, and $T$ is the time ordering operator. For
the vector-potential gauge, along the same line, from
(\ref{eq:density_fkt6_1}), (\ref{eq:density_fkt8_1}),
(\ref{eq:density_fkt13_1}), and (\ref{eq:density_fkt14_1}) we can
arrive at the matrix elements
\begin{equation}\label{eq:density_fkt16_1}
    \<q|\hat{{\mathcal R}}_v(t)|q'\>\, =\, T\,e^{\frac{i}{\hbar} \hat{J}_v(t)}\,\<q|\hat{{\mathcal
    R}}(0)|q'\>\,,
\end{equation}
where the operator $\hat{J}_v(t) := (1/M) \int_0^t
d\tau\,p_c(\tau)\,\hat{{\mathcal V}}_{qq'}(t-\tau)$.

Now we simplify the above expressions for $\<q|\hat{{\mathcal
R}}_s(t)|q'\>$ and $\<q|\hat{{\mathcal R}}_v(t)|q'\>$. The operator
$\hat{J}_s(t)$ immediately reduces to $i
\hbar\,\<\hat{q}(t)\>_s\,(\partial_q +
\partial_{q'}) + \<\hat{p}(t)\>_s\,(q-q')$ from the
well-known exact expression
\begin{equation}\label{eq:linear_response1}
    \textstyle\<{\hat{\mathcal O}}(t)\>_s - \<{\hat{\mathcal O}}(0)\>\, =\, \int_0^t d\tau F(\tau)\,\chi_{{\mathcal
    O}q}(t-\tau)\,,
\end{equation}
where $\hat{{\mathcal O}} \in \{\hat{q}, \hat{p}\}$, obtained
directly from the linear response theory \cite{ING98} (note here
that $\<\hat{q}(0)\> = \<\hat{q}\>_{\beta} = 0$ and $\<\hat{p}(0)\>
= \<\hat{p}\>_{\beta} = 0$ \cite{WEI08,KIM10}, and $\<\hat{p}(t)\>_s
= M \<\dot{\hat{q}}(t)\>_s$). Due to the fact that $[q-q',
\partial_q +
\partial_{q'}] = 0$, equation (\ref{eq:density_fkt16}) then becomes
\begin{equation}\label{eq:density_fkt19}
    {\textstyle\<q|\hat{{\mathcal R}}_s(t)|q'\>\, =\, e^{\frac{i}{\hbar}
    \<\hat{p}(t)\>_s\,(q-q')}}\,
    \sum_{n=0}^{\infty} \textstyle\frac{\left(-\<\hat{q}(t)\>_s\right)^n}{n!}\left(\partial_q +
    \partial_{q'}\right)^n\,\<q|\hat{{\mathcal R}}(0)|q'\>\,.
\end{equation}
Let $y := q + q'$ so that $\partial_q +
\partial_{q'} = 2 \partial_y$. Then we can easily obtain
\begin{equation}\label{eq:density_fkt20}
    \textstyle\left(\partial_q + \partial_{q'}\right)^n\,\<q|\hat{{\mathcal R}}_s(0)|q'\>\,
    =\, 2^n \left(\frac{\partial}{\partial y}\right)^n \frac{1}{\sqrt{2\pi \<\hat{q}^2\>_{\beta}}}\,
    \exp\left(-\frac{y^2}{8\,\<\hat{q}^2\>_{\beta}} - \frac{\<\hat{p}^2\>_{\beta}\,(q - q')^2}{2 \hbar^2}\right)\,,
\end{equation}
which can subsequently be expressed in terms of the Hermite
polynomial as
\begin{equation}\label{eq:density_fkt21}
    \textstyle\frac{1}{\sqrt{2\pi \<\hat{q}^2\>_{\beta}}} \left(\frac{-1}{\sqrt{2
    \<\hat{q}^2\>_{\beta}}}\right)^n e^{-\frac{\<\hat{p}^2\>_{\beta}}{2 \hbar^2}\,(q - q')^2} e^{-z^2}\,H_n(z)
\end{equation}
with $z = y/\sqrt{8 \<\hat{q}^2\>_{\beta}}$. Here we used the
identity $H_n(z) = (-1)^n\,e^{z^2} (d/dz)^n\,e^{-z^2}$ \cite{ABS74}.
Then, with the aid of the identity $e^{2 z t - t^2} =
\sum_{n=0}^{\infty} \{H_n(z)/n!\}\,t^n$ \cite{ABS74}, equation
(\ref{eq:density_fkt19}) finally reduces to the exact expression
\begin{equation}\label{eq:density_fkt25}
    \textstyle\<q|\hat{{\mathcal R}}_s(t)|q'\>\, =\, \frac{1}{\sqrt{2\pi
    \<\hat{q}^2\>_{\beta}}}\,\exp\left\{-\frac{\<\hat{p}(t)\>_s^2}{2\,\<\hat{p}^2\>_{\beta}}\right\}\,\exp\left\{\frac{-1}{2\,\<\hat{q}^2\>_{\beta}}
    \left(\frac{q + q'}{2} - \<\hat{q}(t)\>_s\right)^2\,-\,\frac{\<\hat{p}^2\>_{\beta}}{2 \hbar^2} \left(q - q' + \frac{\hbar}{i}
    \frac{\<\hat{p}(t)\>_s}{\<\hat{p}^2\>_{\beta}}\right)^2\right\}\,.
\end{equation}
The normalization $\text{Tr}\,\hat{\mathcal R}_s(t) = 1$ can easily
be shown with the aid of \cite{ABS74}
\begin{equation}\label{eq:interal_of_exponential1}
    \textstyle\int_{-\infty}^{\infty} dq\,e^{-(a q^2 + 2 b\,q)}\, =\, \sqrt{\frac{\pi}{a}}\,e^{b^2/a}\,.
\end{equation}
For the vector-potential gauge, along the same line, after making
some calculations leading to (\ref{eq:density_fkt19}) with
$\hat{J}_v(t)$ in place of $\hat{J}_s(t)$, we can obtain
\begin{equation}\label{eq:density_fkt19_1}
    {\textstyle\<q|\hat{{\mathcal R}}_v(t)|q'\>\, =\, e^{\frac{i}{\hbar} \<\hat{p}(t)\>_v\,(q-q')}}\,
    \sum_{n=0}^{\infty} \textstyle\frac{\left(-\<\hat{q}(t)\>_v\right)^n}{n!}\left(\partial_q +
    \partial_{q'}\right)^n\,\<q|\hat{{\mathcal R}}(0)|q'\>\,,
\end{equation}
where we identified $\<\hat{q}(t)\>_v$ and $\<\hat{p}(t)\>_v$,
respectively, as $\int_0^{t} d\tau\,p_c(\tau)
\dot{\chi}_{qq}(t-\tau)$ and $\int_0^{t} d\tau\,p_c(\tau)
\dot{\chi}_{pq}(t-\tau)$ [note that $\chi_{sr}(t) = - M
\dot{\chi}_{qq}(t)$ and $\chi_{sp}(t) = - M \dot{\chi}_{pq}(t)$],
verified directly from the linear response theory with
(\ref{eq:density_fkt2_1}) in place of (\ref{eq:density_fkt2}). In
fact, with the aid of integration by parts it can easily be shown
that
\begin{equation}\label{eq:scalar_pot-vector_pot1}
    \<\hat{q}(t)\>_s\,=\,\<\hat{q}(t)\>_v\;\; ;\;\;
    \<\hat{p}(t)\>_s \,=\,\<\hat{p}(t)\>_v\,+\,p_c(t)\,.
\end{equation}
Consequently we can immediately obtain
\begin{equation}\label{eq:density_fkt26}
    \<q|\hat{{\mathcal R}}_v(t)|q'\>\, =\, \left.\<q|\hat{{\mathcal R}}_s(t)|q'\>\right|_{\<\hat{p}(t)\>_s \to
    \<\hat{p}(t)\>_v}\, =\, e^{-\frac{i}{\hbar} p_c(t) (q - q')}\,\<q|\hat{{\mathcal R}}_s(t)|q'\>\,.
\end{equation}
Obviously, we have $\text{Tr}\,\hat{\mathcal R}_v(t) = 1$ from
(\ref{eq:density_fkt25}).

It is instructive now to consider $\<\hat{q}^2(t)\>$ and
$\<\hat{p}^2(t)\>$ in both scalar-potential and the vector-potential
gauges. Using (\ref{eq:density_fkt25}) we can easily obtain
\begin{subequations}
\begin{eqnarray}
    \<\hat{q}^2(t)\>_s &=& \<\hat{q}^2\>_{\beta}\,+\,\<\hat{q}(t)\>_s^2\label{eq:scalar_pot_correlation1}\\
    \<\hat{p}^2(t)\>_s &=& \<\hat{p}^2\>_{\beta}\,+\,\<\hat{p}(t)\>_s^2\,.\label{eq:scalar_pot_correlation2}
\end{eqnarray}
\end{subequations}
Similarly, we can also have
\begin{subequations}
\begin{eqnarray}
    \<\hat{q}^2(t)\>_v &=& \text{Tr} \left\{\hat{q}^2\,\hat{{\mathcal
    R}}_v(t)\right\}\, =\, \<\hat{q}^2(t)\>_s\\
    \<\hat{p}^2(t)\>_v &=&
    \<\hat{p}^2\>_{\beta}\,+\,\<\hat{p}(t)\>_v^2\,.\label{eq:vector_pot_correlation2}
\end{eqnarray}
\end{subequations}
As a result, the instantaneous internal energy in the
scalar-potential gauge
\begin{equation}\label{eq:internal_energy1}
    \<\hat{H}_s(t)\>_s\, =\, \frac{\<\hat{p}^2(t)\>_s}{2 M}\, +\,
    \frac{M}{2} \omega_0^2\,\<\hat{q}^2(t)\>_s
\end{equation}
is not necessarily identical to its vector-potential gauge
counterpart, namely,
\begin{equation}\label{eq:internal_energy2}
    \<\hat{H}_s(t)\>_s\,-\,\<\hat{H}_s(t)\>_v\,=\,\frac{1}{2 M}\,\left\{\<\hat{p}(t)\>_s^2 -
    \<\hat{p}(t)\>_v^2\right\}\,\ne\,0\,.
\end{equation}
At first glance, it looks like a paradox. However, we have a rather
simple justification for this \cite{ELB87}: In the scalar-potential
gauge problem, the experiment is performed in such a way that we
turn on an external field at $t=0$ and then turn off at $t=t_f$.
Afterwards we measure the fluctuation $\<\hat{p}^2(t_f)\>_s$. In the
vector-potential-gauge setting, on the other hand, we need to turn
off the vector potential $p_c(t_f) = \int_0^{t_f} F(\tau)\,d\tau$
rather than the external field. Consequently the fluctuation
$\<\hat{p}^2(t_f)\>_v$ differs from its scalar-potential counterpart
in such a way that
\begin{equation}
    \<\hat{p}^2(t_f)\>_v\, =\, \<\hat{p}^2\>_{\beta}\,+\,\{\<\hat{p}(t_f)\>_s -
    p_c(t_f)\}^2\,,
\end{equation}
which is actually accordance with the result in
(\ref{eq:vector_pot_correlation2}) with
(\ref{eq:scalar_pot-vector_pot1}). It is, however, in general
physically unrealistic to carry out an experiment in which the
vector potential is turned off.

Comments deserve here. First, it is interesting to note a
time-independent behavior of the purity measure
\begin{equation}
    \text{Tr}\,\hat{{\mathcal R}}_s^2(t)\, =\, \text{Tr}\,\hat{{\mathcal R}}_v^2(t)\, =\,
    \frac{\hbar}{2 \sqrt{\<\hat{q}^2\>_{\beta}\,\<\hat{p}^2\>_{\beta}}}\,,
\end{equation}
obtained directly from (\ref{eq:density_fkt25}) and
(\ref{eq:density_fkt26}) with (\ref{eq:interal_of_exponential1}),
respectively.

Secondly, as was shortly pointed out in Sect.
\ref{sec:introduction}, whereas the equilibrium quantities
$\<\hat{q}^2\>_{\beta}$ and $\<\hat{p}^2\>_{\beta}$ in equations
(\ref{eq:x_correlation1}) and (\ref{eq:x_dot_correlation1}) can
exactly be obtained from the scheme of linear response theory, this
is not the case for their non-equilibrium counterparts
$\<\hat{q}^2(t)\>$ and $\<\hat{p}^2(t)\>$; in fact, by using
(\ref{eq:density_fkt6}) we can arrive at the expression
\begin{equation}\label{eq:beyond_linear_response1}
    \textstyle\<\hat{q}^2(t)\>_s\, =\, \<\hat{q}^2\>_{\beta}\,+\,\int_0^t
    d\tau\,\chi^{(1)}(t-\tau)\,F(\tau)\,+\,\int_0^t d\tau\,F(\tau) \int_0^{\tau} d\tau'\,F(\tau')\,\chi^{(2)}(t-\tau, t-\tau')\,,
\end{equation}
where $\chi^{(1)}(t) = \chi_{q^2 q}(t) = \<[\hat{q}^2(t),
\hat{q}]\>_{\beta} = 0$, and the 2nd-order response function
$\chi^{(2)}(t,\tau) = \<[[\hat{q}^2, \hat{r}(t)],
\hat{r}(\tau)]\>_{\beta}$ can be obtained from the cyclic invariance
of the trace, which subsequently reduces to
$2\,\chi_{qq}(t)\,\chi_{qq}(\tau)$ with the aid of
(\ref{eq:r_time1}). The relation $\int_0^t d\tau \int_0^{\tau}
d\tau' = \int_0^t d\tau' \int_{\tau'}^t d\tau$ \cite{SCH01} then
allows equation (\ref{eq:beyond_linear_response1}) to become
\begin{equation}\label{eq:beyond_linear_response2}
    \textstyle\<\hat{q}^2(t)\>_s - \<\hat{q}^2\>_{\beta}\, =\, 2 \int_0^t
    d\tau'\,F(\tau')\,\chi_{qq}(t-\tau')\,\int_{\tau'}^{t} d\tau\,F(\tau)\,\chi_{qq}(t-\tau)\,,
\end{equation}
which is also identical to
\begin{equation}\label{eq:beyond_linear_response3}
    \textstyle 2 \int_0^t d\tau'\,F(\tau')\,\chi_{qq}(t-\tau') \int_0^{\tau'} d\tau\,F(\tau)\,\chi_{qq}(t-\tau)
\end{equation}
directly resulting from (\ref{eq:beyond_linear_response1}) with
exchange of the two integral variables $\tau$ and $\tau'$. From
(\ref{eq:linear_response1}), (\ref{eq:beyond_linear_response2}) and
(\ref{eq:beyond_linear_response3}) we can immediately recover the
exact result in (\ref{eq:scalar_pot_correlation1}). Similarly we can
do the same job for $\<\hat{p}^2(t)\>_s$ and then for their
vector-potential gauge counterparts, respectively.

Subsequently, we can also obtain, with the aid of
(\ref{eq:interal_of_exponential1}), the higher-order fluctuations
such as
\begin{subequations}
\begin{eqnarray}
    \<\hat{q}^3(t)\>_s &=& \textstyle\<\hat{q}(t)\>_s
    \left\{3\,\<\hat{q}^2\>_{\beta}\,+\,\<\hat{q}(t)\>_s^2\right\}\label{eq:higher_order_fluctuation1}\\
    \<\hat{p}^3(t)\>_s &=& \textstyle\<\hat{p}(t)\>_s
    \left\{3\,\<\hat{p}^2\>_{\beta}\,+\,\<\hat{p}(t)\>_s^2\right\}\label{eq:higher_order_fluctuation2}
\end{eqnarray}
\end{subequations}
and
\begin{subequations}
\begin{eqnarray}
    \hspace*{-1.1cm}&&\<\hat{q}^4(t)\>_s\, =\, \textstyle 3\,\<\hat{q}^2\>_{\beta}^2\,+\,6\,\<\hat{q}^2\>_{\beta}\,\<\hat{q}(t)\>_s^2\,+\,\<\hat{q}(t)\>_s^4\\
    \hspace*{-1.1cm}&&\<\hat{p}^4(t)\>_s\, =\, \textstyle 3\,\<\hat{p}^2\>_{\beta}^2\,+\,6\,\<\hat{p}^2\>_{\beta}\,\<\hat{p}(t)\>_s^2\,+\,\<\hat{p}(t)\>_s^4\,,
\end{eqnarray}
\end{subequations}
etc. Their vector-potential counterparts immediately appear with the
replacement of $\<\hat{p}(t)\>_s \to \<\hat{p}(t)\>_v$.

\section{Numerical Analysis within the Drude damping model}\label{sec:numerical}
We carry out the numerical analysis in the scheme of the well-known
Drude model (with a cut-off frequency $\omega_d$ and a damping
parameter $\gamma_o$), which is a prototype for physically realistic
damping. It is then known that \cite{KIM07}
\begin{subequations}
\begin{eqnarray}
    \hspace*{-.5cm}\<\hat{q}^2\>_{\beta}^{(d)} &=& {\textstyle \frac{1}{M}}
    \sum_{l=1}^3 {\textstyle \lambda_d^{(l)}\,\left\{\frac{1}{\beta \underline{\omega_l}}\,
    +\, \frac{\hbar}{\pi}\; \psi\left(\frac{\beta \hbar \underline{\omega_l}}{2 \pi}\right)\right\}}\label{eq:x_drude}\\
    \hspace*{-.5cm}\<\hat{p}^2\>_{\beta}^{(d)} &=& {\textstyle -M}
    \sum_{l=1}^3 {\textstyle \lambda_d^{(l)}\,\underline{\omega_l}^2\,\left\{\frac{1}{\beta \underline{\omega_l}}\,
    +\, \frac{\hbar}{\pi}\; \psi\left(\frac{\beta \hbar \underline{\omega_l}}{2 \pi}\right)\right\}}\label{eq:p_drude}
\end{eqnarray}
\end{subequations}
in terms of the digamma function $\psi(y) = d\,\ln\Gamma(y)/dy$
\cite{ABS74}, where $\underline{\omega_1} = \Omega$,
$\underline{\omega_2} = z_1$, $\underline{\omega_3} = z_2$, and the
coefficients
\begin{equation}\label{eq:coefficients}
    \textstyle\lambda_d^{(1)}\; =\; \frac{z_1\,+\,z_2}{(\Omega\,-\,z_1) (z_2\,-\,\Omega)}\; ;\;
    \lambda_d^{(2)}\; =\; \frac{\Omega\,+\,z_2}{(z_1\,-\,\Omega)
    (z_2\,-\,z_1)}\; ;\; \lambda_d^{(3)}\; =\; \frac{\Omega\,+\,z_1}{(z_2\,-\,\Omega) (z_1\,-\,z_2)}\,.
\end{equation}
Here we have employed, in place of $(\omega_0, \omega_d, \gamma_o)$,
the parameters $({\mathbf w}_0, \Omega, \gamma)$ through the
relations \cite{FOR06}
\begin{equation}\label{eq:parameter_change0}
    \textstyle\omega_0^2\, :=\, {\mathbf w}_0^2\; \frac{\Omega}{\Omega\, +\, \gamma}\; ;\;
    \omega_d\, :=\, \Omega\, +\, \gamma\; ;\;
    \gamma_o\, :=\, \gamma\, \frac{\Omega\, (\Omega\, +\, \gamma)\,
    +\, {\mathbf w}_0^2}{(\Omega\, +\, \gamma)^2}\,,
\end{equation}
and then $z_1 = \gamma/2 + i {\mathbf w}_1$ and $z_2 = \gamma/2 - i
{\mathbf w}_1$ with ${\mathbf w}_1 = \sqrt{({\mathbf w}_0)^2 -
(\gamma/2)^2}$. For the underdamped case $({\mathbf w}_0 \geq
\gamma/2)$ we have $z_2 = \bar{z}_1$ whereas $z_1,\,z_2 > 0$ for the
overdamped case $({\mathbf w}_0 < \gamma/2)$. The susceptibility in
the Drude damping model is also well-known as \cite{FOR06,KIM06}
\begin{equation}\label{eq:susceptibility_drude1}
    \tilde{\chi}_{qq}^{(d)}(\omega)\, =\, -\frac{1}{M} \frac{\omega\,+\,i (\Omega + z_1 + z_2)}{(\omega\,+\,i \Omega)\,(\omega\,+\,i z_1)\,(\omega\,+\,i
    z_2)}\,.
\end{equation}
With the aid of (\ref{eq:inverse_laplace}) we can easily obtain the
response function
\begin{equation}\label{eq:response_fkt_drude1}
    \chi_{qq}^{(d)}(t)\, =\, -\frac{1}{M} \frac{\left(z_1^2 - z_2^2\right)\,e^{-\Omega\,t}\,+\,\left(z_2^2 -
    \Omega^2\right)\,e^{-z_1\,t}\,+\,\left(\Omega^2 - z_1^2\right)\,e^{-z_2\,t}}{\left(\Omega -
    z_1\right)\,\left(z_1 - z_2\right)\,\left(z_2 - \Omega\right)}\,.
\end{equation}
This is real-valued and holds true for both underdamped and
overdamped cases.

We first apply a static external field $F_1(t) = A_1\,\Theta(t)$
(d.c. field) and then an oscillatory field $F_2(t) = A_2\,\sin
\omega_f t$ (a.c. field). By substituting equation
(\ref{eq:response_fkt_drude1}) to (\ref{eq:linear_response1}) we can
easily obtain both $\<\hat{q}(t)\>_s^{(d)}$ and
$\<\hat{p}(t)\>_s^{(d)}$ within the Drude damping model in closed
form for the two different forms of external force, respectively.
Similarly we can also have their vector-potential counterparts in
closed form. Figures \ref{fig:fig1}-\ref{fig:fig4} demonstrate
temporal behaviors of $\<\hat{q}^n(t)\>_s^{(d)}$ with $n = 1,2$ for
different damping and control parameters. Further, it is instructive
to study a temporal behavior of a distance between the initial
equilibrium state $\hat{{\mathcal R}}(0)$ and the non-equilibrium
state $\hat{{\mathcal R}}(t)$. To do so, we adopt a well-defined
measure $D^2(t) = \text{Tr}\,(\{\hat{{\mathcal R}}(t) -
\hat{{\mathcal R}}(0)\}^2)$, introduced in \cite{GRA98}, which is,
independent of the dimension of the Liouville space, between 0 and
2. With the aid of (\ref{eq:interal_of_exponential1}) we can then
have
\begin{equation}\label{eq:density_operator_distance1}
    \textstyle D_s^2(t)\,=\,\frac{\hbar}{\sqrt{\<\hat{q}^2\>_{\beta}\,\<\hat{p}^2\>_{\beta}}}
    \left(1 - \exp\left\{-\frac{1}{4} \left(\frac{\<\hat{q}(t)\>_s^2}{\<\hat{q}^2\>_{\beta}} +
    \frac{\<\hat{p}(t)\>_s^2}{\<\hat{p}^2\>_{\beta}}\right)\right\}\right)
\end{equation}
and $D_v^2(t) = D_s^2(t)|_{\<\hat{p}(t)\>_s \to \<\hat{p}(t)\>_v}$.
In figure \ref{fig:fig5} this measure within the Drude damping model
is demonstrated for different external fields and temperatures.

\section{Concluding remarks}\label{sec:conclusions}
In summary, we have discussed the field-induced dynamics in the
scheme of quantum Brownian oscillator at an arbitrary temperature.
We have then derived the reduced density operator in closed form of
the coupled oscillator in a non-equilibrium state at an arbitrary
time. In doing so, we have applied both scalar-potential and
vector-potential gauges for the interaction Hamiltonian. We believe
that this exact expression for the reduced density operator will
provide a useful starting point, e.g., for later useful discussions
of quantum thermodynamics and quantum information theory within
quantum Browian oscillator.

The author would like to thank G. Mahler for helpful discussions. He
also appreciates all constructive remarks of an anonymous referee.
%

%
%
\vspace*{0.5cm}

Figure~\ref{fig:fig1}: $y = \<\hat{q}(t)\>_s^{(d)}$ versus time $t$
for a static field $F_1(t) = A_1\,\Theta(t)$. The response
$\<\hat{q}(t)\>_s^{(d)}$ is temperature-independent; refer to
equations (\ref{eq:linear_response1}) and
(\ref{eq:response_fkt_drude1}). Here ${\mathbf w}_0 = \Omega = M =
1$. (1) dash ($\gamma = 5$, overdamped): From bottom to top, (black,
$A_1 = 1$), (violet, $A_1 = 1.5$), and (red, $A_1 = 2$); (2) solid
($\gamma = 0.1$, underdamped): From bottom to top, (blue, $A_1 =
1$), (maroon, $A_1 = 1.5$), and (green, $A_1 = 2$).\vspace*{0.3cm}

Figure~\ref{fig:fig2}: $y = \<\hat{q}(t)\>_s^{(d)}$ versus time $t$
for an oscillatory field $F_2(t) = A_2\,\sin \omega_f t$. The
response $\<\hat{q}(t)\>_s^{(d)}$ is temperature-independent. Here
${\mathbf w}_0 = \Omega = M = A_2 = 1$; (1) violet dot: $\gamma =
0.1$ (underdamped) and $\omega_f = 1 \approx \omega_0$, resonant
[{\em cf}. (\ref{eq:parameter_change0})]; (2) blue dashdot: $\gamma
= 0.1$ (underdamped) and $\omega_f = 1.5$; (3) black dash: $\gamma =
5$ (overdamped) and $\omega_f = 1$; (4) red solid: $\gamma = 5$
(overdamped) and $\omega_f = 1.5$.\vspace*{0.3cm}

Figure~\ref{fig:fig3}: $y = \<\hat{q}^2(t)\>_s^{(d)}$ versus time
$t$ for $F_1(t) = A_1\,\Theta(t)$. For $\<\hat{q}^2(t)\>_s^{(d)}$
refer to equations (\ref{eq:scalar_pot_correlation1}) and
(\ref{eq:x_drude}). Here $\hbar = k_B = {\mathbf w}_0 = \Omega = M =
1$. (1) black dashdot ($\gamma = 5$, overdamped, and $A_1 = 1$):
From bottom to top, dimensionless temperature $k_B T/\hbar {\mathbf
w}_0 = 0.01, 2, 5$; (2) blue dash ($\gamma = 0.1$, underdamped, and
$A_1 = 1$): From bottom to top, $k_B T/\hbar {\mathbf w}_0 = 0.01,
2, 5$; (3) red solid ($\gamma = 0.1$, underdamped, and $A_1 = 2$):
From bottom to top, $k_B T/\hbar {\mathbf w}_0 = 0.01, 2,
5$.\vspace*{0.3cm}

Figure~\ref{fig:fig4}: $y = \<\hat{q}^2(t)\>_s^{(d)}$ versus time
$t$ for $F_2(t) = A_2\,\sin \omega_f t$. Here $\hbar = k_B =
{\mathbf w}_0 = \Omega = M = A_2 = 1$. (1) black dot ($\gamma = 5$,
overdamped, and $\omega_f = 1$): From bottom to top, $k_B T/\hbar
{\mathbf w}_0 = 0.01, 2, 5$; (2) blue solid ($\gamma = 0.1$,
underdamped, and $\omega_f = 1$, resonant): From bottom to top, $k_B
T/\hbar {\mathbf w}_0 = 0.01, 2, 5$; (3) red dash ($\gamma = 0.1$,
underdamped, and $\omega_f = 1.5$): From bottom to top, $k_B T/\hbar
{\mathbf w}_0 = 0.01, 2, 5$.\vspace*{0.3cm}

Figure~\ref{fig:fig5}: $y = D_s^2(t)$ versus time $t$ within the
Drude damping model. For $D_s^2(t)$ refer to
(\ref{eq:density_operator_distance1}). Here $\hbar = k_B = {\mathbf
w}_0 = \Omega = M = A_1 = A_2 = 1$, and $\gamma = 0.1$, underdamped.
(1) dash (for d.c. field): From top to bottom, dimensionless
temperature $k_B T/\hbar {\mathbf w}_0 = 0.01, 2, 5$; (2) solid (for
a.c. field): the same as for dash, with $\omega_f = 1 \approx
\omega_0$, resonant.
%
%
\onecolumn{
\begin{figure}[htb]
\centering\hspace*{-1.75cm}{\includegraphics[scale=0.9]{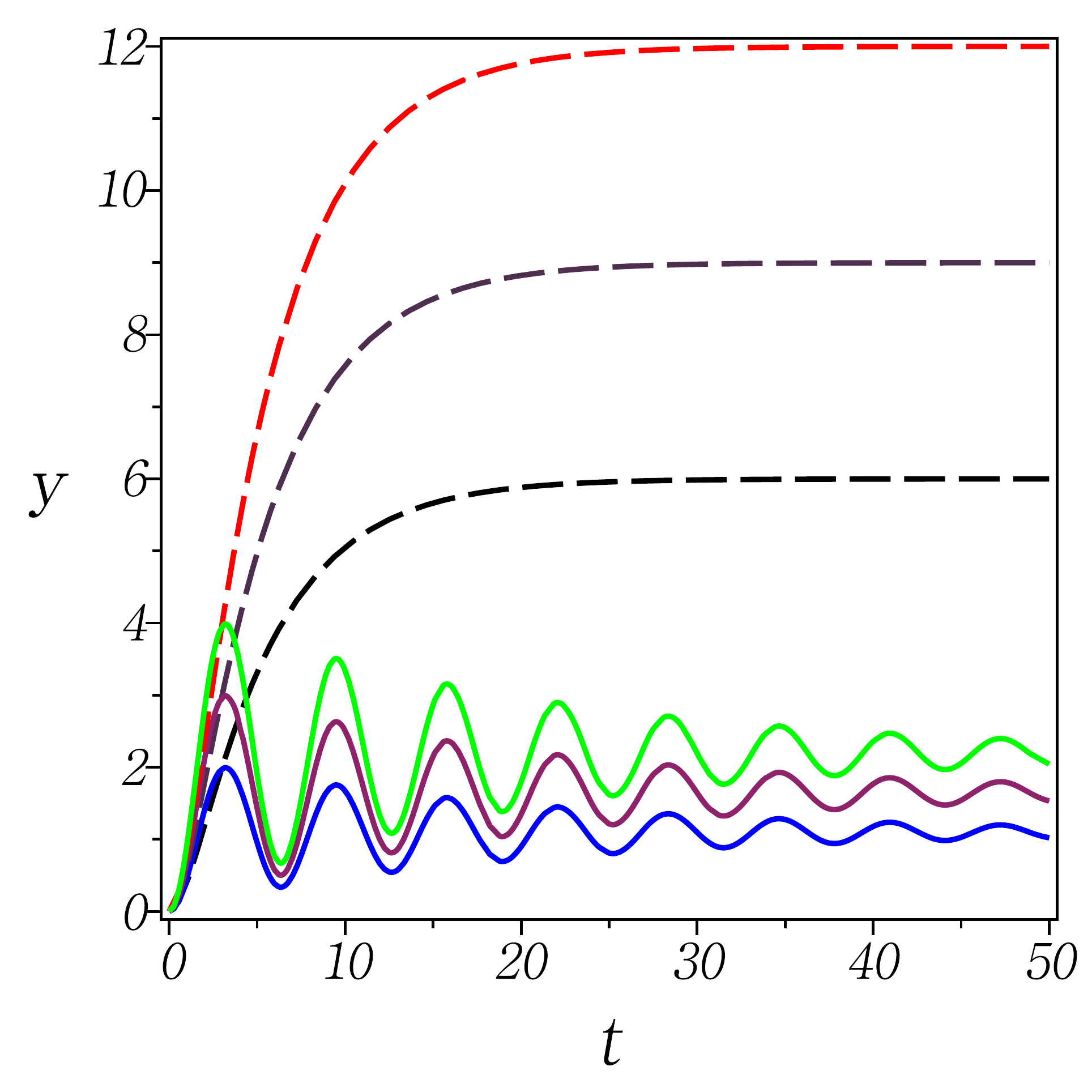}
\caption{\label{fig:fig1}}}
\end{figure}
%
%
\begin{figure}[htb]
\centering\hspace*{-1.75cm}{\includegraphics[scale=0.9]{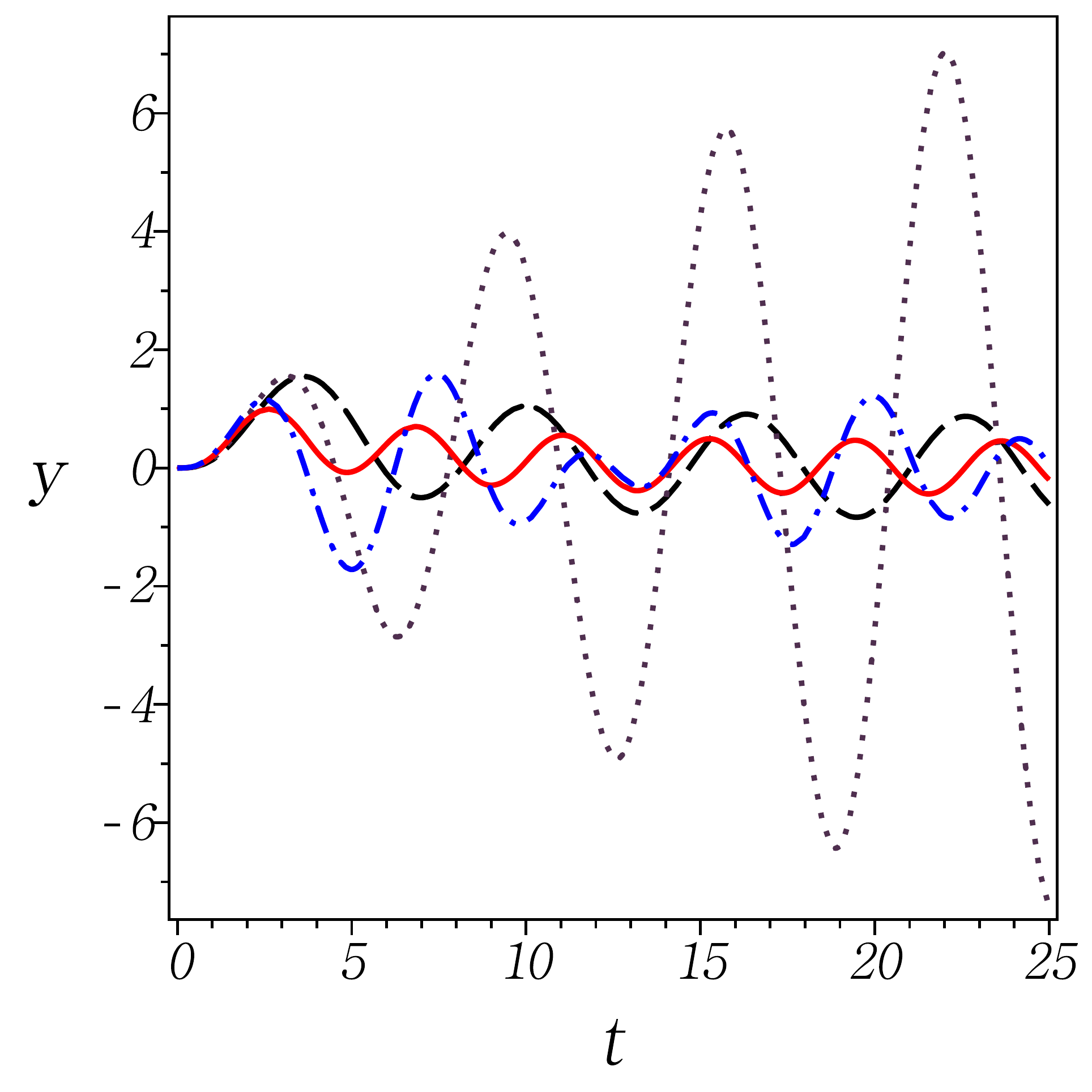}
\caption{\label{fig:fig2}}}
\end{figure}
%
%
\begin{figure}[htb]
\centering\hspace*{-1.75cm}{\includegraphics[scale=0.9]{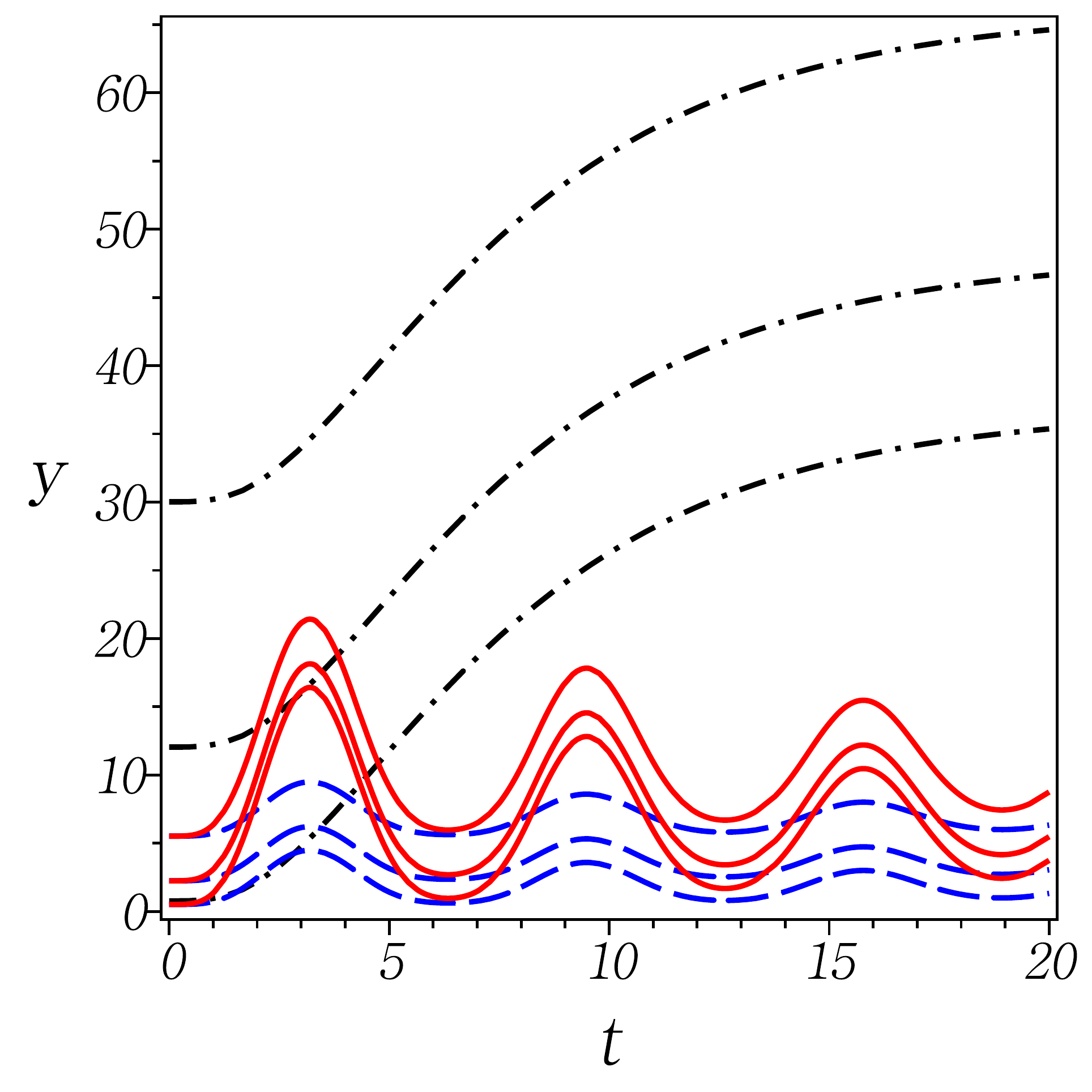}
\caption{\label{fig:fig3}}}
\end{figure}
%
%
\begin{figure}[htb]
\centering\hspace*{-1.75cm}{\includegraphics[scale=0.9]{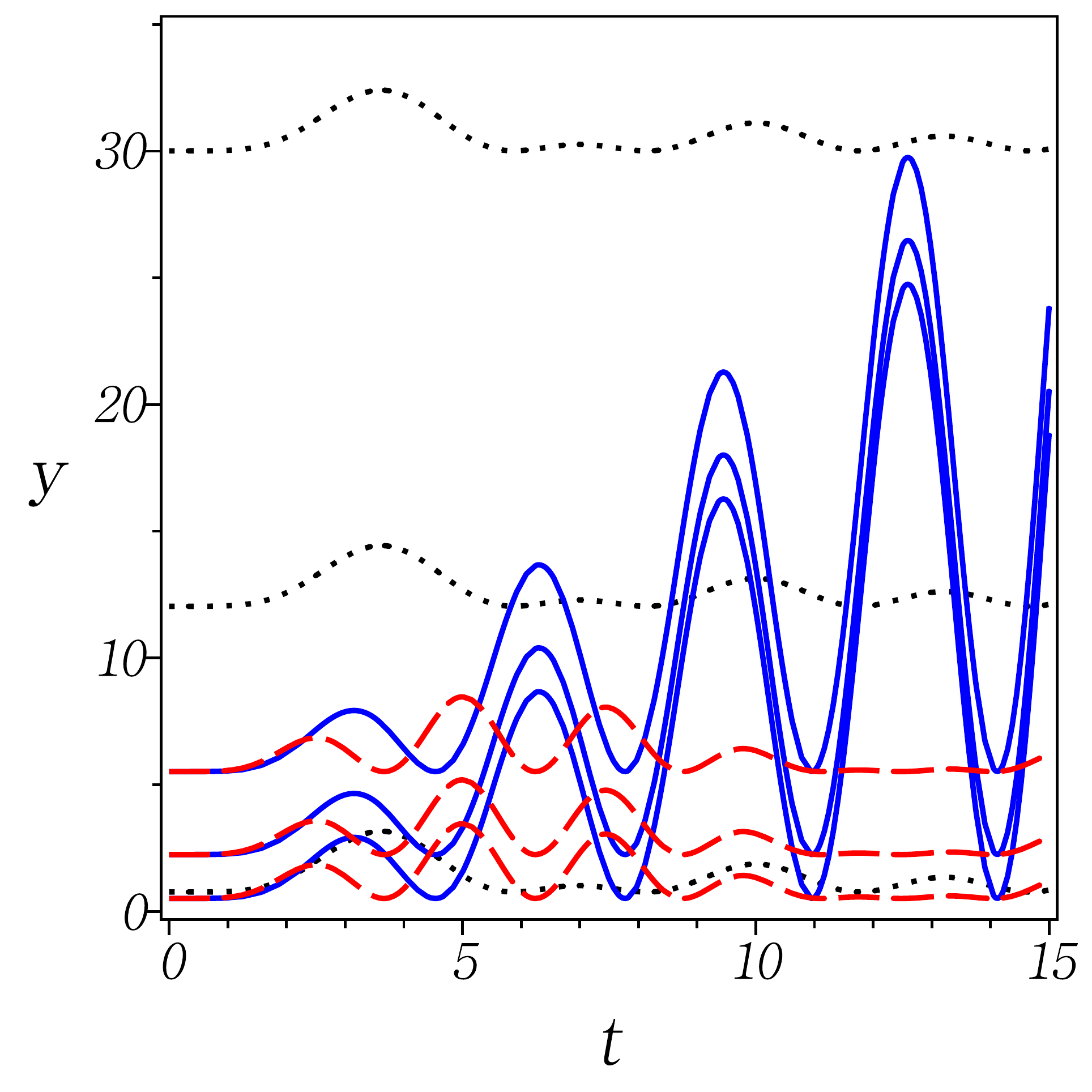}
\caption{\label{fig:fig4}}}
\end{figure}
%
%
\begin{figure}[htb]
\centering\hspace*{-1.75cm}{\includegraphics[scale=0.9]{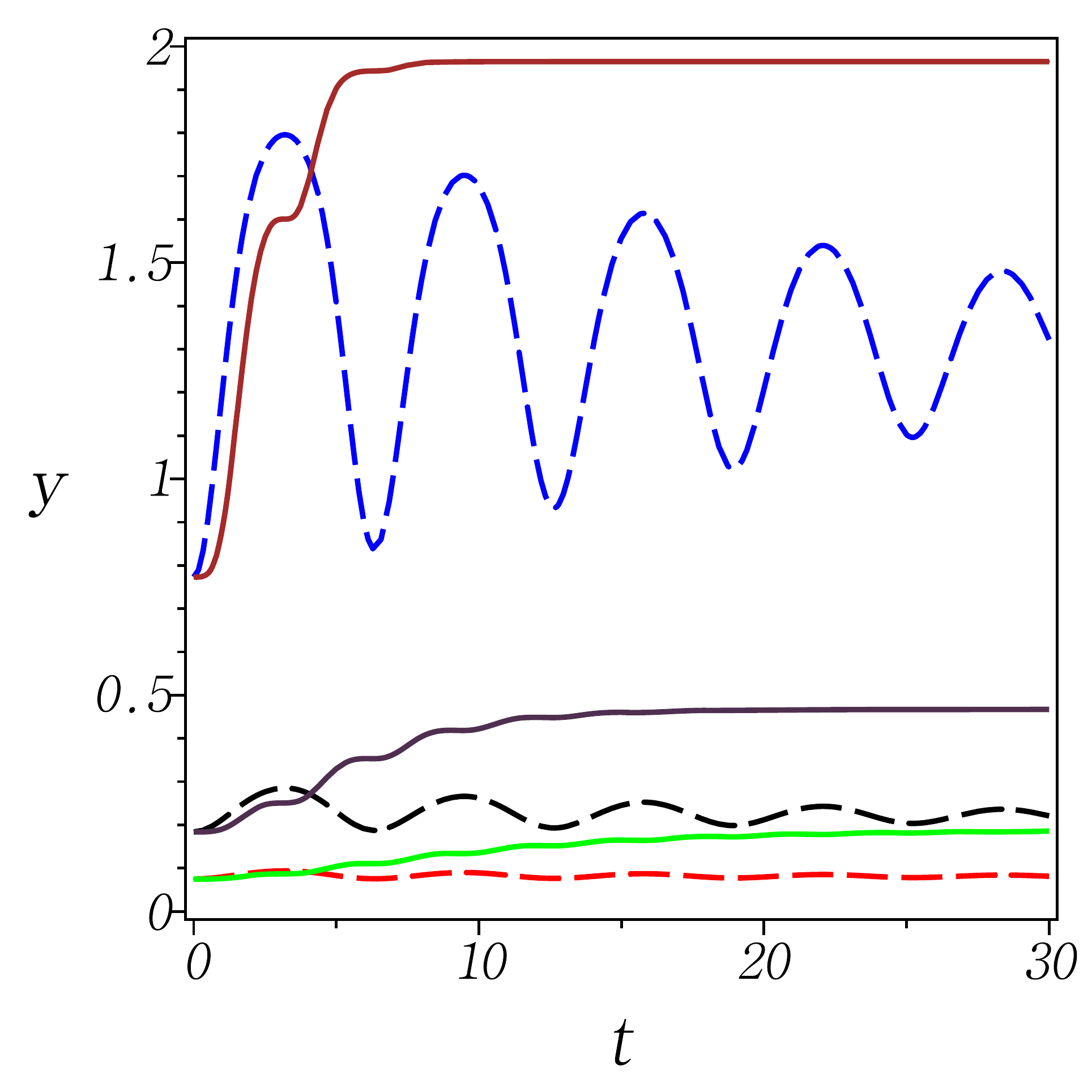}
\caption{\label{fig:fig5}}}
\end{figure}
}

\begin{thebibliography}{1}
\bibitem{ILK10} Non-linear reponses in exactly solvable
systems have already been discussed in different contexts, e.g.,
Ref. \cite{CUG07} in the context of the statistical mechanics for
glassy systems; the spherical model treated therein may also be
viewed in an extended sense as a collection of quantum harmonic
oscillators.
\bibitem{ILK07} In \cite{KIM07}, the (classical) noise correlation
$\<\hat{\xi}(t)\,\hat{\xi}(t')\>_{\rho_{b'}} = M \gamma(t-t')/\beta$
is given just after equation (8), which should be replaced by
equation (\ref{eq:qm_noise_correlation1}) of the current paper for
the quantum case. This minor correction is, obiously, completely
irrelevant to all message of Ref. \cite{KIM07}, though.
\end{thebibliography}
\end{document}